# Topological phase transition in MoTe$_2$: A Review

Suvodeep Paul[*], Saswata Talukdar, Ravi Shankar Singh, and Surajit Saha[*]

*Department of Physics, Indian Institute of Science Education and Research Bhopal, 462066, India*

([*] Corresponding authors: suvodeeppaul100@gmail.com, surajit@iiserb.ac.in)

**Abstract**

Transition metal dichalcogenides (TMDs) are a branch of two-dimensional materials which in addition to having an "easy-to-exfoliate" layered structure, also host semiconducting, metallic, superconducting, and topological properties in various polymorphs with potential applications. MoTe$_2$ is an example of such a TMD, which shows semiconducting (in 2$H$ phase), metallic (in 1$T'$ phase), topological Weyl semimetallic and superconducting behavior (in $T_d$ phase). Consequently, an extensive amount of research has been done on MoTe$_2$, particularly on the topological phase transition between the metallic-type 1$T'$ phase and the topological $T_d$ phase. This phase transition has been reviewed and its association with the crystal structure, charge transport, and electronic band structure is elaborately discussed. Also, the effect of various stimuli like reduced dimensionality, pressure, charge doping, and chemical substitution, which affect the structural transition as well as the superconducting transition temperatures is reviewed; thereby, suggesting certain correlations between the apparently unrelated structural and superconducting phase transitions. The review also brings out some open questions which are likely to interest the community to address the physics associated with the phase transition and its potential applications.

## 1. Introduction

The term 'topology'[1] is associated with the branch of mathematics that deals with the properties of geometric shapes that are preserved under the effect of continuous deformations like stretching and bending. As a result, a doughnut and a coffee mug are topologically equivalent, implying that the two shapes can be interchangeably obtained by means of continuous deformations without the requirement of tearing or gluing new portions. In 2016, David J. Thouless, F. Duncan M. Haldane, and J. Michael Kosterlitz were awarded the Nobel prize in Physics for their contributions in using topological theories to quantum subatomic particles in order to predict exotic states of matter. Various systems hosting such exotic topological behavior have been reported in the past couple of decades and promise an excellent platform for the revelation of new physics with a plethora of exciting applications.

Before we proceed to discuss about various topological systems that have been widely explored recently, we may consider a typical example of a 2D film of electrons exposed to a magnetic field perpendicular to the film (as shown in Figure 1a), to understand the importance of topology in quantum matter. The electrons in the bulk of the film undergo revolutions in circular orbits about the magnetic field direction, and therefore, conduct no electricity. However, near the edge of the film, the orbits are open and connected, resulting in a unidirectional flow of electrons along the edges. Figure 1b shows that the edge conductivity of this material would not be affected by small deformations or defects, making this edge state a platform to realize perfect electron transport. The robustness of the edge state of such a material is attributed to its topological character.

Physical examples of such topological character were observed in graphene which shows the integer quantum Hall effect[2]. In presence of a magnetic field, the electrons start moving in circular orbits with the cyclotron frequency $\omega_c$, which gives rise to quantized Landau levels with energies, $\varepsilon_m = \hbar\omega_c (m+1/2)$. When 'N' Landau levels are filled, the Hall conductivity is given by: $\sigma_{xy} = Ne^2/h$, where e (electronic charge) and h (Planck's constant) are universal constants. The Hall conductivity has been reported to a precision of 1 part in $10^9$ by Klitzing *et al.*[3]. The Quantum Hall effect is explained by invoking a topological phase because certain fundamental properties like the quantized value of the Hall conductance and the number of gapless edge channels, show excellent robustness to smooth changes in the material parameters[1]. The alteration of these properties would require the topological invariants to change, which can only be possible if the material is passed through a topological phase transition. Though the first examples of topological character in physical systems were observed in 2D systems like graphene, the topological materials known today have also expanded to 3D materials, which include topological insulators (TIs)[1] and Weyl semimetals (WSMs)[4].

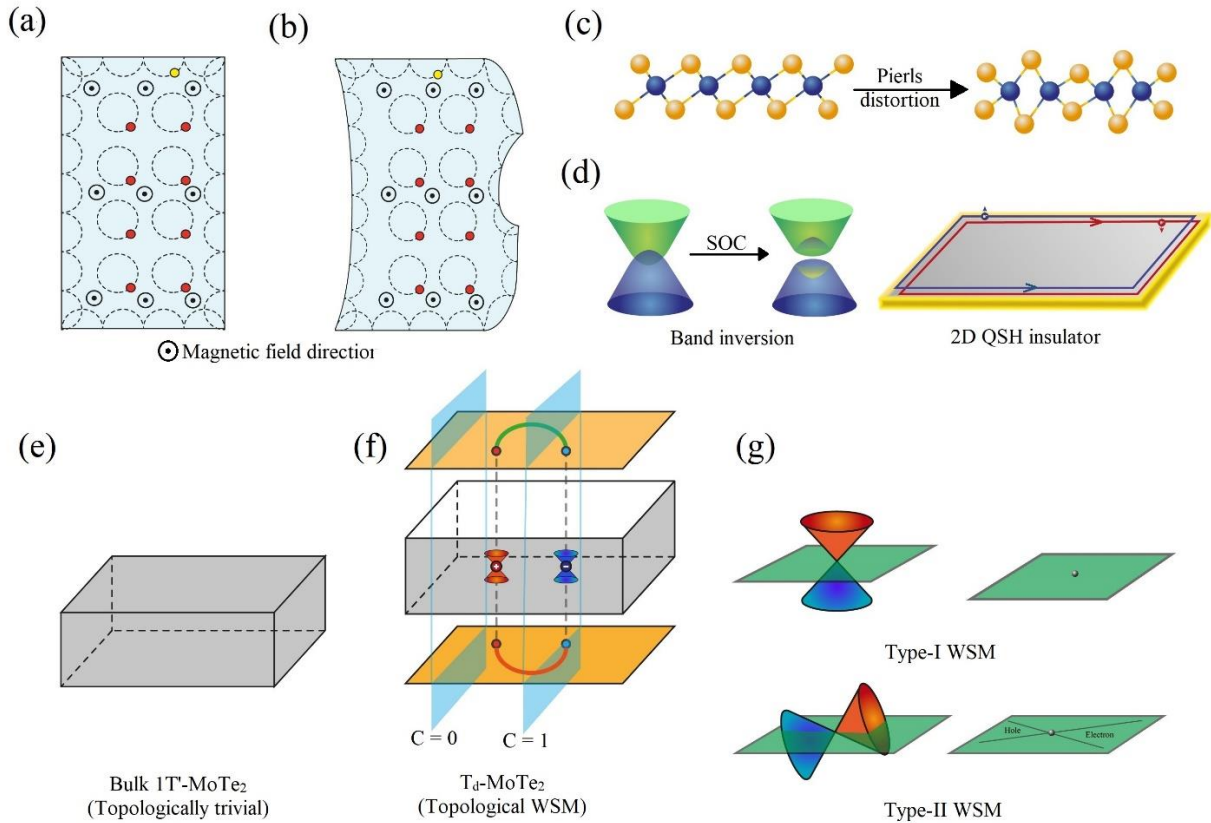

**Figure 1. a)** A classical 2D system of electrons exposed to a perpendicular magnetic field, resulting in rotation of electrons (represented by red circles) about the magnetic field directions. The electron at the edge (represented by yellow circle) follows an incomplete and connected orbit such that it undergoes a unidirectional motion, constituting an edge current. **b)** Deformations in the 2D system of electrons are observed to have no effect on the unidirectional flow of edge electrons because of a topological protection. **c)** The Pierls distortion in the $1T$ phase, resulting in the formation of the $1T'$ structure. **d)** The distortion in the $1T$ phase to form the $1T'$ structure is associated with a band inversion phenomenon. The spin-orbit coupling in the system results in the opening of a spin Hall band-gap, thereby, allowing Quantum spin Hall effect in the monolayer $1T'$ crystal. **e)** The bulk $1T'$ crystal, which is formed by stacking the $1T'$ layers in a monoclinic structure, shows topologically non-trivial character. **f)** The breaking of inversion symmetry in the bulk $1T'$ crystal gives rise to the $T_d$ phase, which is characterized by Weyl points (WPs) of +1 and -1 chirality. When projected on a surface, the WPs of opposite chirality act as starting and ending points of open arcs called Fermi arcs (represented by green and red lines). The blue planes cutting through the material are assigned topological invariants C=1 and C=0, based on whether the surfaces are chosen in between the WPs of opposite chirality or not. **g)** The Weyl semimetals may be of Type-I or Type-II. While the Type-I Weyl semimetals (WSMs) are represented by Fermi surfaces that are reduced to a point (touching point of the conduction and valence bands), the Type-II WSMs show electron and hole pockets at the Fermi surface separated by the WP.

Transition metal dichalcogenides (TMDs) are a class of layered materials represented by the molecular formula $MX_2$ (M = transition metal, X = chalcogen) that hosts a wide variety of unique properties[5–8]. The layers of these TMDs constitute a chain of transition metal atoms sandwiched between two chains of chalcogen atoms. Within this sandwich structure, the M and X atoms may, however, exhibit different atomic arrangements giving rise to various polymorphs. The $2H$ phase constitutes trigonal prismatic structure with the M atoms located at the center of the prism. The $1T$ phase shows octahedral coordination of the M atoms to the X atoms. The $2H$ phase has been predominantly studied because of its thickness-dependent semiconducting properties that promise excellent applications in nanoelectronics and optoelectronics devices. The $1T$ phase, on the other hand, exhibits metallic character[9]. Recent investigations reported a slight distortion in the central M chain of the $1T$ phase, thereby, giving rise to the $1T'$ structure. This distortion, also known as the Pierls distortion (Figure 1c), has been reported to result in a band inversion phenomenon in the $1T'$ phase[10], which is followed by a band-gap opening induced by spin-orbit coupling, to give rise to Quantum spin Hall (QSH) insulating phase in monolayers of $1T'$ TMDs[11] (Figure 1d). The QSH phase is a topological phase that is characterized by an insulating 2D plane and conducting edge states. $MoTe_2$, which belongs to the TMD family, is of particular significance, because the $1T$ phase in $MoTe_2$ is energetically unstable and, therefore, it spontaneously undergoes a Pierls distortion to attain the $1T'$ structure[12]. The QSH materials show a segregation of spins (instead of charges in normal Hall effect) induced by strong spin-orbit coupling (SOC) in the material (instead of an externally applied magnetic field in normal Hall effect), resulting in a spin current. Consequently, the QSH materials have great potential in spintronic applications. Unfortunately, spintronic applications are limited to very low temperatures due to the small QSH band-gaps in most QSH materials[13–15]. Once again, $1T'$-$MoTe_2$ proves useful as a significantly higher band-gap opening of ~60 meV was reported for few-layered flakes of $MoTe_2$[12]. Bulk $1T'$-$MoTe_2$ essentially comprises of $1T'$ monolayers held together by weak van der Waals forces. In the bulk, the $1T'$ layers stack in a monoclinic structure which possesses inversion symmetry (which is absent in the case of monolayer $1T'$). The bulk $1T'$ phase, on the other hand, has no topological character and has been reported to show normal metallic behavior[16] (Figure 1e). However, recent theoretical studies indicate that the bulk $1T'$ phase may also be a higher order topological insulator, but there are no experimental evidences to this end[17]. Interestingly, bulk $1T'$-$MoTe_2$ undergoes a structural phase transition below ~250 K to give rise to the $T_d$ phase[18], which breaks the inversion symmetry (as in the case of monolayer $1T'$). Various theoretical reports have predicted the $T_d$ phase of $MoTe_2$ (and $WTe_2$) to be a Type-II WSM[19,20], as elaborated under.

WSMs are condensed matter systems where Weyl fermions are observed as low energy excitations near the WPs. In 1929, German mathematician and theoretical physicist Hermann Weyl derived the Weyl equation that could describe the motion of massless relativistic particles[21]. While there has been no experimental evidence of Weyl fermions in high energy particle physics, various condensed matter systems (WSMs) have been reported to show a linear band structure in the vicinity of band touching points (WPs) such that the band structure mimics the Weyl equation[22–24]. Therefore, the electrons near the WPs are expected to host properties like Weyl fermions, but at non-relativistic speeds. The WSMs are characterized by Weyl cones that touch at WPs in the band structure (similar to Dirac cones observed in graphene) but with a chirality associated with each Weyl point. Therefore, unlike Dirac points, WPs must always exist in pairs with chirality of +1 and -1, respectively[4]. Materials that preserve both time reversal and lattice inversion symmetry cannot host Weyl fermions because the states of opposite chirality would annihilate each other, rendering a normal state[25]. Therefore, it is the intrinsic requirement for WSMs that either the time reversal symmetry or the lattice inversion symmetry must be broken. A projection of the WPs on the surface gives rise to topologically protected surface states (Fermi arcs), which are essentially open arcs that connect the two WPs having opposite chirality (Figure 1f). WSMs are classified into Type-I and Type-II

(Figure 1g). In the Type-I WSMs, the Weyl cones touch at the Weyl point which results in the Fermi surface to be reduced to a point. The Type-II WSMs show an inclination of the Weyl cones such that the Fermi surface consists of electron and hole pockets, which touch at the WPs. Unlike the Type-I WSMs, the Type-II WSMs do not respect the Lorentz symmetry, which is a fundamental requirement of high energy particle physics. This further enhances the excitement of studying the Type-II WSMs as they might host novel properties, which are not predicted for normal Weyl fermions. The $T_d$ phase of MoTe$_2$ (and WTe$_2$) breaks the lattice inversion symmetry of the $1T'$ crystal, to achieve the topological Type-II WSM phase[20]. Consequently, the phase transition from the $1T'$ to the $T_d$ phase is a topological phase transition, which is associated with: 1) a structural transition involving the breaking of inversion symmetry, 2) renormalization of the electronic band structure which changes the charge conduction properties of the material, and 3) appearance of topologically protected trivial surface states and non-trivial Fermi arcs joining pair of WPs[4]. Based on the various changes associated with the phase transition mentioned above, a variety of experimental tools have been used for its detection. While neutron scattering and Raman measurements bring out the structural changes and the breaking of inversion symmetry during the phase transition, charge transport measurements show the renormalization of electronic properties. The topological character of the low temperature $T_d$ phase is brought out by angle resolved photoemission spectroscopy (ARPES) and scanning tunneling microscopy/spectroscopy (STM/STS) measurements, which involve the detection of the topologically protected surface states including Fermi arcs.

We have already discussed about the importance of the centrosymmetric $1T'$ phase and the topological Type-II Weyl semimetallic $T_d$ phase of MoTe$_2$, and the subsequent phase transition between them. We may note here that in addition to the topological phases, semimetallic MoTe$_2$ also has a superconducting ground state at very low temperatures (T$_c$ ~0.1 K)[16,26–31]. There have been extensive efforts by various groups over the past few years to enhance the superconducting transition temperature. Substantial enhancement in the superconducting T$_c$ has been induced by various perturbations which include the effect of reduced dimensionality[32,33], pressure[16], chemical substitution[26,34–36] etc. However, little attention has been paid towards the effect of these perturbations on the topological phase transition temperature (T$_s$) between the $1T'$ and $T_d$ phases. A careful observation of the evolution of the T$_s$ and T$_c$ of MoTe$_2$ induced by various perturbations mentioned above suggests that the two apparently unrelated phenomena, *viz.*, the topological phase transition and the superconducting ground state, may have a correlation. This correlation between the topological and the superconducting phase transitions of MoTe$_2$, if present, may be a new avenue to look for by the theoreticians and the experimentalists alike, in the already rich field of topological materials.

In this review, we will give a detailed description of various physical properties associated with the phase transition between $1T'$ and $T_d$-MoTe$_2$. In section 2, we will discuss the structural phase transition and the associated breaking of inversion symmetry along with the related experimental evidences. In section 3, we will discuss about the electronic properties associated with the phase transition and the resulting effect on the charge transport. In section 4, we will discuss the topological nature of the transition which is manifested by the appearance of open Fermi arcs on the surface. In section 5, we will discuss about the various perturbative effects and their influences on the topological phase transition in MoTe$_2$. This will include effects of reduced dimensionality, charge doping, pressure, and chemical substitution. In section 6, we will discuss the possibility of a correlation between the superconducting ground state of MoTe$_2$ and the topological phase transition. In section 7, we will pose certain open questions which will encourage future research in this area. Finally, we will summarize the various results reported on MoTe$_2$ in Section 8.

## 2. Topological vs structural phases: Inversion symmetry breaking

The phase transition from the $1T'$ (centrosymmetric) phase to the $T_d$ (non-centrosymmetric) phase is associated with a structural transition involving the breaking of inversion symmetry. The $1T'$ phase (Figure

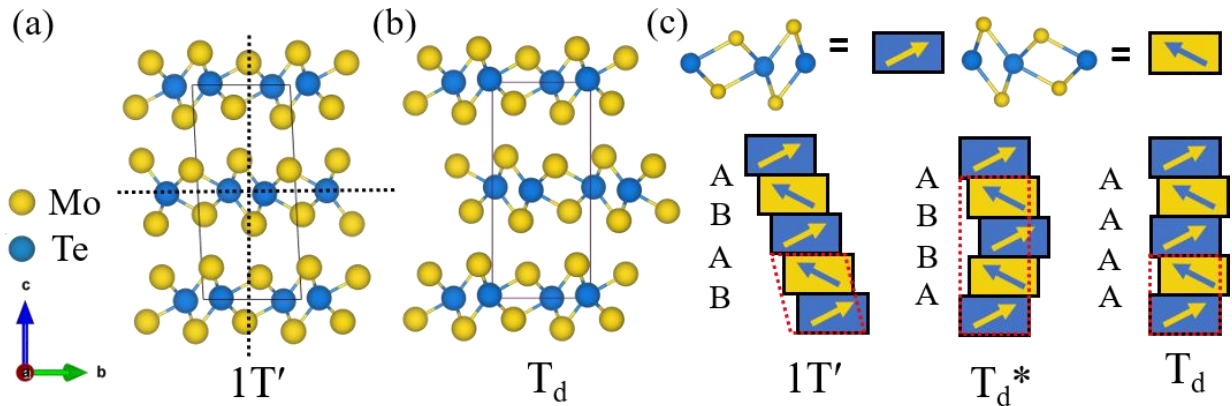

**Figure 2.** Crystal structures of: **a)** $1T'$ and **b)** $T_d$ phases of MoTe$_2$. **c)** Generation of the $1T'$, $T_d^*$, and $T_d$ phases of MoTe$_2$ by stacking yellow and blue building blocks by operations A and B (defined in text). The $1T'$, $T_d^*$, and $T_d$ phases constitute operations A and B in sequence AB, ABBA, and AA, respectively. Part **(c)** is adapted with permission. [38] Copyright 2019, American Physical Scciety.

2a) has a monoclinic structure and belongs to the $P2_1/m$ space group[37]. As shown in Figure 2a, the intersection of the dotted horizontal and vertical lines on the unit cell represents the center of inversion symmetry of the crystal. The $T_d$ phase (Figure 2b), on the other hand has an orthorhombic unit cell and breaks the inversion symmetry[37]. Essentially, the $1T'$ and the $T_d$ phases are obtained by differential stacking of $1T'$ layers of MoTe$_2$. Figure 2c demonstrates how the stacking operations of two types (labelled A and B) can be repeated along the c axis to obtain the $1T'$ and the $T_d$ phases[38]. The A operation maps one layer on to the layer below it (the two layers are represented by the blue and yellow rectangles, respectively) such that the complete $T_d$ stack may be formed by repeating the A operations in AA sequence. The B operation is similar to the A with an additional translation of ±0.15 lattice unit in the in-plane direction, where the + and – signs are effective in alternate layers. The $1T'$ stacking may be represented by a repetition of the A and B operations in AB sequence. It has been recently predicted from neutron scattering measurements that while heating the MoTe$_2$ crystals from the $T_d$ phase to the $1T'$ phase, a new intermediate phase may be formed in the vicinity of the transition, which is represented by the repetition of A and B operations in an ABBA sequence and this intermediate phase is termed as the $T_d^*$ phase[38]. However, the $T_d^*$ phase has not been observed while cooling the MoTe$_2$ crystals from the $1T'$ to the $T_d$ phase.

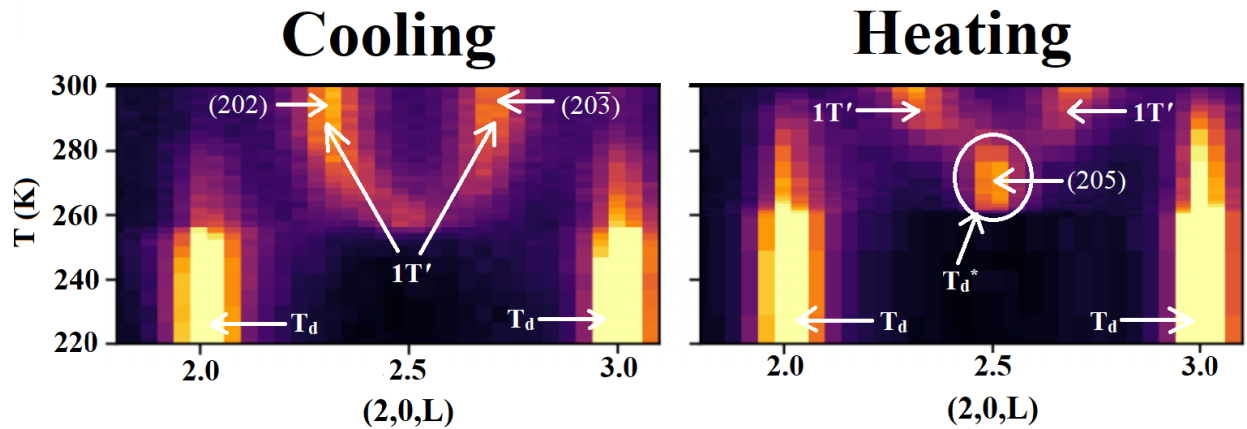

**Figure 3.** Neutron scattering intensity maps as a function of temperature along the (2,0,L) line on cooling and heating. Features corresponding to the Bragg peaks of the $T_d$ and $1T'$ phases of MoTe$_2$ were observed in both cycles, while the feature associated with the $T_d^*$ phase was observed only in the heating cycle. Figure is reproduced with permission.[38] Copyright 2019, American Physical Scciety.

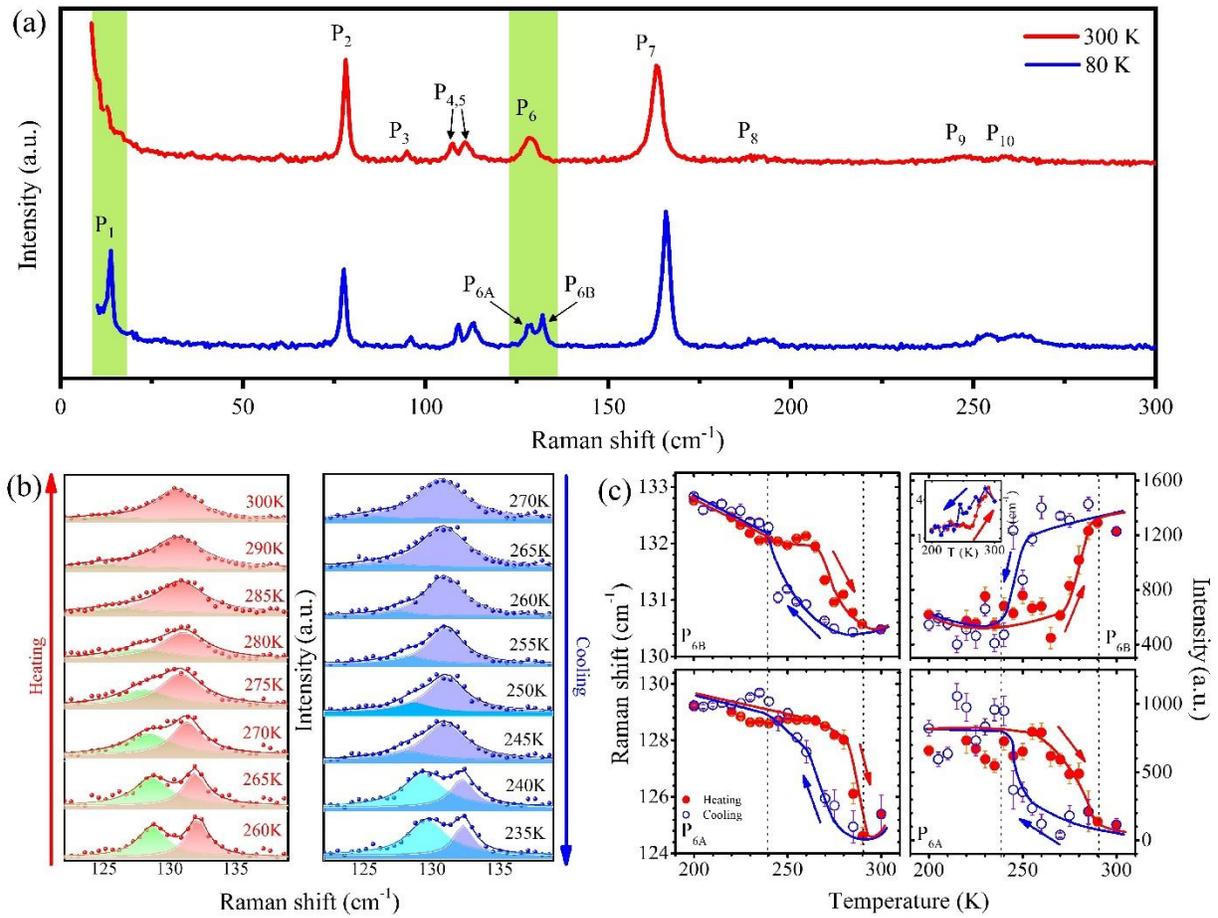

**Figure 4.** Raman detection of the structural phase transition. **a)** The Raman spectra of the $1T'$ (red) and $T_d$ (blue) phases with labeled modes. The green rectangles mark the differences in the Raman signatures of the two phases. The phonon $P_1$ appears and the $P_6$ mode splits into $P_{6A}$ and $P_{6B}$ modes in $T_d$ phase. **b)** Evolution of $P_6$ mode into constituent modes $P_{6A}$ and $P_{6B}$ represent the structural phase transition in MoTe$_2$. **c)** The structural transition is observed as a hysteretic change in phonon properties in the vicinity of the transition temperature, $T_s$. Part (b) and (c) are reproduced with permission.[32] copyright 2022, American Physical Soceity.

There are various reports of neutron scattering experiments in order to study the structural changes in the vicinity of the phase transition temperature, $T_s$[38,39]. Figure 3 shows the intensity maps along (2, 0, L) for various temperatures, as was reported from elastic neutron scattering measurements by Tao *et al.*[38]. The $1T'$ phase is characterized by the (202) and (20$\bar{3}$) Bragg peaks, which appear at L=2.3 and 2.7, respectively, as can be seen in Figure 3. The $T_d$ phase is represented by the (202) and (203) peaks, observed at L=2 and 3, respectively. The appearance (disappearance) of the features associated with the $T_d$ ($1T'$) phase occur at different temperatures in the cooling and heating cycles, clearly revealing a hysteretic behavior. Additionally, a feature is observed at L=2.5 in the heating cycle, which is clearly absent in the cooling data. This feature is attributed to the possible appearance of the $T_d^*$ phase at the vicinity of the phase transition from the $T_d$ to the $1T'$ phase, as discussed above. The exact mechanism of the structural phase transition must be studied in further detail in order to understand the absence of the $T_d^*$ phase in the cooling cycle[38,40].

The structural transition and the consequent breaking of inversion symmetry also result in substantial changes in the phonon eigenvectors, and therefore, may be detected through Raman measurements[32,37,49,41–48]. The unit cells of both the $1T'$ and $T_d$ phases constitute 12 atoms and hence give rise to 36 normal modes of vibration[37]. The corresponding irreducible representations for the phonons at the Brillouin zone of the $1T'$ and the $T_d$ crystals are represented by equations (1) and (2).

$$\Gamma_{bulk,1T'} = 12A_g + 6A_u + 6B_g + 12B_u \quad \ldots (1)$$

$$\Gamma_{bulk,Td} = 12A_1 + 6A_2 + 6B_1 + 12B_2 \quad \ldots (2)$$

In equations (1) and (2), the phonon symmetries of the $1T'$ structure are of $A_g$, $A_u$, $B_g$, and $B_u$ types, all of which represent the presence of inversion symmetry in the crystal. While the symmetries with suffix $g$ represent the symmetric vibrations with respect to the center of inversion symmetry and are Raman active, the symmetries with suffix $u$ represent antisymmetric vibrations with respect to the center of symmetry and are Raman inactive (they are infrared active vibrations). By virtue of the breaking of inversion symmetry, the phonons undergo a renormalization in the $T_d$ phase giving rise to symmetries of $A_1$, $A_2$, $B_1$, and $B_2$, all of which are Raman active[37]. Figure 4a shows the Raman spectra of the $1T'$ phase (obtained at 300 K) and the $T_d$ phase (obtained at 80 K). A comparison between the two spectra clearly reveals the appearance of a new mode, designated as $P_1$ and the splitting of the mode designated as $P_6$ into $P_{6A}$ and $P_{6B}$ in the $T_d$ phase. As discussed above, this is a consequence of the breaking of inversion symmetry which results in renormalization of the phonon eigenstates associated with the lattice. Therefore, in the region near 11 cm$^{-1}$, the $P_1$ phonon converts from the Raman inactive $B_u$ symmetry to the Raman active $A_1$ symmetry in the $T_d$ phase. Similarly, the region near 129 cm$^{-1}$ constitutes two phonon modes of Raman active $A_g$ and Raman inactive $B_u$ symmetry in $1T'$ phase, both of which convert to Raman active $A_1$ symmetry in the $T_d$ phase. The evolution of the $P_1$ and the $P_6$ phonons have been studied by several groups, to detect the phase transition in $1T'$-MoTe$_2$[32].

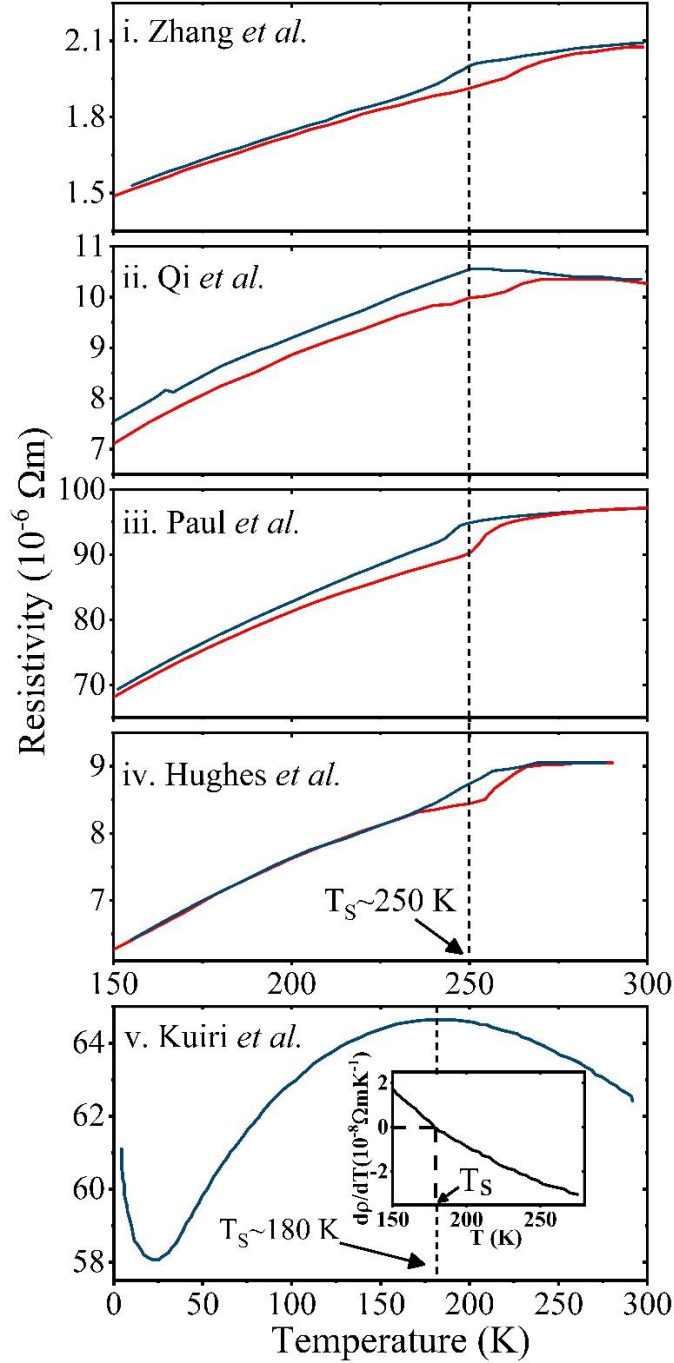

**Figure 5.** Charge transport detection of the phase transition in bulk crystals of MoTe$_2$. The structural phase transition is observed as a hysteretic anomaly (top four panels) in the charge transport behavior near the transition temperature, $T_s$. The bottom panel shows the resistivity data of 11 nm thick flake, which shows the $T_s$ =180 K, where the slope of the resistivity (inset) changes from negative to positive as the temperature is lowered. An upturn in the resistivity below 20 K is also observed at low temperature. Plots **i-v**) are adapted from ref. [16,18,32,37,52], respectively.

Figure 4b shows the Raman spectra during the heating and cooling cycles bringing out the evolution of the $P_6$ mode into the constituents, $P_{6A}$ and $P_{6B}$[32]. The phase transition occurs at different temperatures in the heating and cooling cycles, resulting in a hysteretic behavior. Figure 4c shows the hysteresis observed in various parameters (phonon energies, linewidths, and intensities) of the phonons in the vicinity of the phase transition[32].

## 3. Electronic properties in the vicinity of the phase transition

The transition between the $1T'$ and $T_d$ phases have been known since long due to reports of anomalous charge transport behavior around 250 K[18,50]. The anomaly in the charge transport measurements appeared with a hysteresis and was, therefore, attributed to a first order phase transition. In accordance with temperature dependent X-ray diffraction data[51], the observed anomaly was attributed to a structural phase transition. While the topological nature of the low temperature phase ($T_d$) was only realized very recently[19,20], the hysteretic charge transport behavior continues to be considered as one of the most important techniques to realize the presence of the phase transition. There are many reports on charge transport measurements by various groups revealing the phase transition, some of which are summarized in Figure 5[16,18,35,37,52]. The phase transition temperature $T_s$ is measured from the center of the hysteresis loop, and it shows excellent agreement ($T_s$ ~250 K) across reports from various groups over the years for bulk crystals of MoTe$_2$ (top four panels of Figure 5).

The bottom panel of Figure 5 shows the resistivity data of a 11 nm thick film of MoTe$_2$. It may be noted that the resistivity shows a change in slope (d$\rho$/dT) around ~180 K from negative to positive (shown in the inset of Figure 5) as the sample is cooled. The $1T'$ ($T_d$) phase is characterized by d$\rho$/dT <0 (>0) and the change in the sign of the d$\rho$/dT corresponds to the phase transition temperature, $T_s$. The measured $T_s$ (~180 K) is way lower than the bulk transition temperature (~250 K). This effect will be further elaborated in sections 5.1 and 6. For the bulk flakes (top four panels of Figure 5), it may be noted that the slope of the resistivity (d$\rho$/dT) is negative (Qi *et al.*[16]) or almost zero in the high temperature $1T'$ phase suggesting semiconducting/semimetallic transport as also seen in the resistivity measured for thinner flakes of MoTe$_2$ (Kuiri *et al.*[52]). The high temperature resistivity (when T >$T_s$) is reported to show Arrhenius-type behavior with an activation energy ~4 meV. At low temperatures (in the vicinity of ~20 K), the resistivity shows an upturn which resembles a metal to insulator transition. Kuiri *et al.*[52] confirmed through 1/f noise measurements that the upturn in resistivity is not associated with a metal-insulator transition. They have proposed a model incorporating contributions from electron-electron and electron-phonon interactions to explain the low-temperature behavior of the resistivity. At T >20 K, the resistivity essentially varies linearly with temperature due to electron-phonon interactions, as was also previously reported in graphene[52,53]. At T <20K, the resistivity shows a 1/T dependence, which is theoretically predicted in WSMs due to electron-electron interactions near the Weyl nodes[52,54].

## 4. Topological nature of the phase transition

Based on the literature available, it is clear that the transition between the $1T'$ and the $T_d$ phases is brought about by several changes which includes changes in structural, electronic properties, and topological aspects, which occur simultaneously. Therefore, probing either of these features ensures the presence of the intended topological transition. However, in order to observe the topological nature of the phase transition, we need to delve deep into the study of the electronic band structure and the Fermi surfaces of MoTe$_2$. As the $T_d$ phase is reached by cooling the $1T'$ crystal below ~250 K, we expect the appearance of WSM features. This would mean that there can be pairs of WPs in the band structure and appearance of Fermi arcs connecting these pairs of WPs. Also, by virtue of the Type-II WSM nature of the $T_d$ phase, the WPs are expected to be located at touching points of electron and hole pockets (Figure 1g). In order to observe these features, various groups have performed angle resolved photoemission spectroscopy (ARPES) on MoTe$_2$[43,55–58]. Accordingly, band structure calculations have revealed the presence of electron and hole pockets at the Fermi level, E$_F$. Figure 6a shows the calculated spectral function in the $k_x$-$k_y$ plane at E$_F$ by Deng et al.[55]. The bell-shaped contour away from the Γ-point represents the electron pocket, while the bowtie shaped contour near the zone center represents the hole pocket. These features are very nicely captured in the intensity maps at E$_F$ obtained from ARPES measurements at temperature below 20 K (Figure 6b). In fact, two pairs of WPs (W1 and W2) were observed at the touching points of the electron and hole pockets in the calculated spectral function at 0.005 eV and 0.045 eV above E$_F$, respectively (Figure 6c). The topologically non-trivial open Fermi arcs appear from the W1 and W2 Weyl points, as shown by the yellow dashed lines in Figure 6c, and the red lines in the schematic shown in Figure 6d. The Fermi arcs were also clearly observed as intense features in the intensity maps around E$_F$ obtained from ARPES (Figure 6e). The thermal broadening induced at higher temperatures prevent any ARPES measurements in the high-temperature 1T′ phase, which appears above T$_s$ ~250 K in bulk MoTe$_2$. Thirupathaiah et al.[59] performed ARPES measurements at 130 K temperature and observed no appreciable changes in the electronic band structure with respect to measurements performed at low (20 K) temperature.

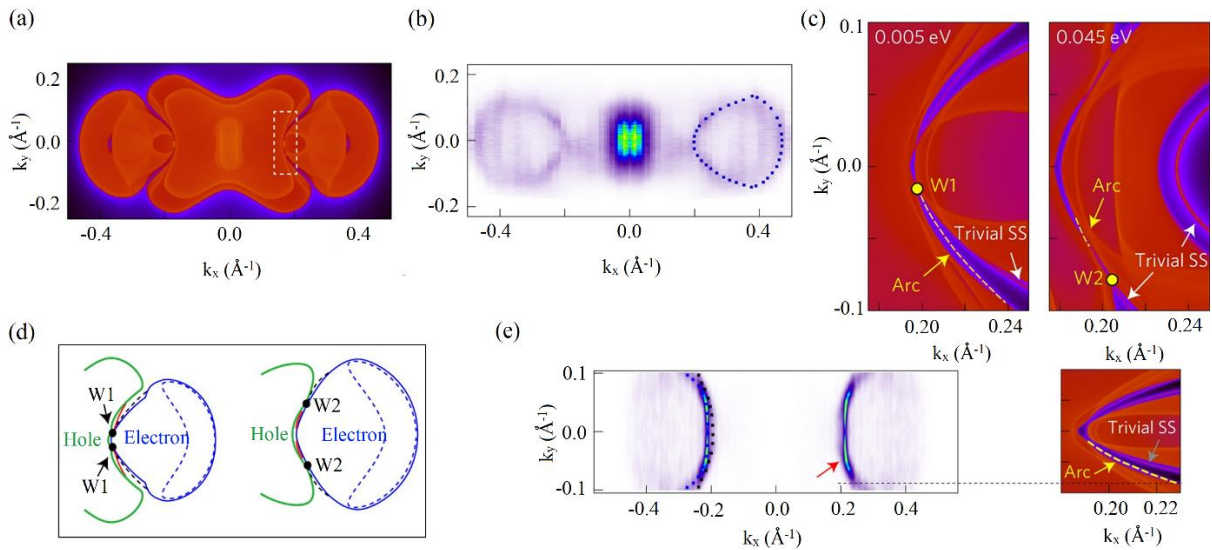

**Figure 6.** Angle-resolved photoemission spectroscopy (ARPES) detection of the phase transition. **a)** Calculated spectral function at E$_F$. **b)** Intensity map measured at E$_F$ with p polarization perpendicular to b axis. The electron and hole pockets are highlighted by blue and green color. **c)** Calculated spectral function at 0.005 eV and 0.045 eV, showing the WPs W1 and W2. **d)** Schematic showing the different electronic structures at the WPs. **e)** ARPES intensity maps at energy E$_F$. The rectangular region shown in **(a)** is zoomed in the right panel of (e) showing the arcs from the topological surface states. Yellow arrows point to the topological surface states and grey arrows point to the trivial surface states. Figure is reproduced with permission.[55] Copyright 2016, Springer Nature.

Berger et al.[60] reported quasiparticle interference (QPI) studies through scanning tunneling microscopy to achieve further insight about the topologically non-trivial surface states in the $T_d$ phase, and their absence in the high temperature 1T′ phase. Similar to prior reports, Berger et al.[60] showed prominent topologically non-trivial surface states from their ARPES measurements performed at low temperature and *ab initio* calculations (Figure 7a) for the $T_d$ phase. They have further used STM imaging to record the differential conductance in real space, which provides important information about the band structure and surface states. The point defects in the crystal act as elastic scatterers resulting in quasiparticle interference patterns based on the momentum transfer in the scattering processes. Figure 7b shows the Fourier transform of the real space differential conductance map, which shows wing-like features in the horizontal direction. The theoretically calculated QPI pattern, shown in Figure 7c, constitutes contributions from scattering events between surface-surface states (red contour), bulk-bulk states (blue contour), and surface-bulk states (green contour). The obtained QPI pattern from STM experiments matches best with the calculated pattern for surface-surface scattering events (red contour). Therefore, we may conclude that the QPI pattern generated from the differential conductance data is largely due to the scattering between surface-surface states. All these observations were made for experiments performed on the $T_d$ phase (at low temperature). Similar experiments performed at room temperature show no features in the QPI, revealing that the surface states are absent in the $1T'$ phase (Figure 7d). However, it may be noted that the thermal broadening at room temperature can be sufficiently large to obscure any signatures from the surface states. Therefore, a differential conductance mapping of the $T_d$ phase was again performed at low temperature but with a bias oscillation of 30 meV to simulate the thermal broadening of 300 K temperature. It was reported that the QPI of the $T_d$ phase shows the wing-like features associated with surface-surface scattering events even with the simulated thermal broadening (Figure 7e). Therefore, it may be concluded that the absence of any signatures in the QPI obtained for $1T'$-MoTe$_2$ is indeed because of the absence of topological surface states.

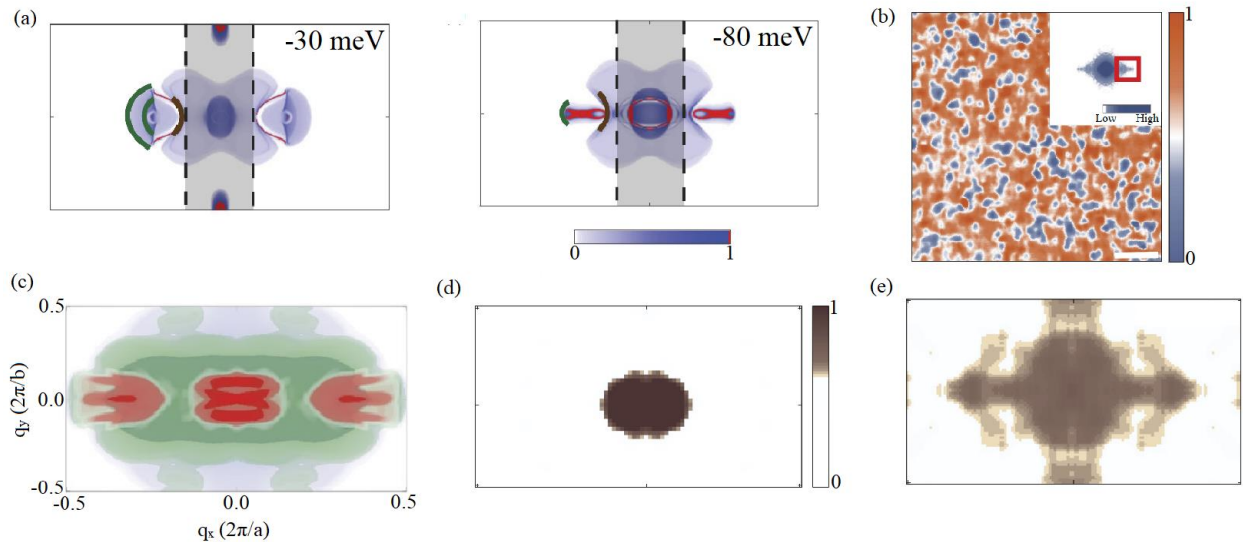

**Figure 7. a)** Energy contours measured through ARPES at various energies. The red color represents the most intense features which correspond to surface states. **b)** Differential conductance (dI/dV) map with inset showing Fourier transform of the real space maps. The Fourier transform inset (quasi particle interference or QPI pattern) shows wings aligned with the a-axis of the crystal (red box). **c)** All three QPI components overlaid, with bulk shown in blue, joint shown in green, and surface, shown in red. **d)** Fourier transform of differential conductance map taken at room temperature showing no features in the QPI. **e)** Fourier transform of the differential conductance map taken at low temperature with a bias oscillation to simulate room temperature thermal broadening. The horizontal wing feature is still visible in the QPI, even in the presence of energy broadening. Figure is adapted from ref. [60].

## 5. Controlling the phase transition through perturbations

Different structural phases of MoTe$_2$ and the corresponding topological phase transition between them have been already discussed in detail. We may note here that in addition to the topological phase transition, MoTe$_2$ exhibits another exceptional property, *viz.*, its superconducting ground state. $T_d$-MoTe$_2$ has been reported to show superconductivity at very low temperature (T$_c$ = 0.1 K)[16,26–31]. Naturally, there have been numerous efforts to enhance the superconducting T$_c$, which have resulted in very significant enhancements. The enhancements in superconducting T$_c$ have been induced by various perturbative influences like the effect of reduced dimensionality, pressure, chemical substitution etc. Though it has been observed that the above-mentioned perturbations to the system have led to the change in the topological phase transition between the $1T'$ and $T_d$ phases, the topic which has not received much of an attention yet. We will dedicate this section to the understanding of the response in the topological phase transition because of the perturbative effects that have been introduced to improve the superconducting $T_c$ of MoTe$_2$.

### 5.1. Effect of reduced dimensionality and charge doping

One of the great benefits of layered materials like TMDs is the possibility to obtain dangling bonds-free thin flakes of these materials by exfoliation process. The weak out-of-plane van der Waals interactions that hold the layers of these materials together allow the separation of atomically thin pristine layers, which often show drastically different properties compared to their bulk counterparts. Further, the separation of atomically thin flakes can also be useful for the fabrication of flexible nanoscale devices, which is one of the most promising applications expected from 2D materials. The thinning down of MoTe$_2$ to ultrathin regimes has also been reported to show enhancement in the superconducting $T_c$[29]. We will review various reports on thickness-dependent studies on MoTe$_2$ to observe and understand the effect of reduced dimensionality on the topological phase transition.

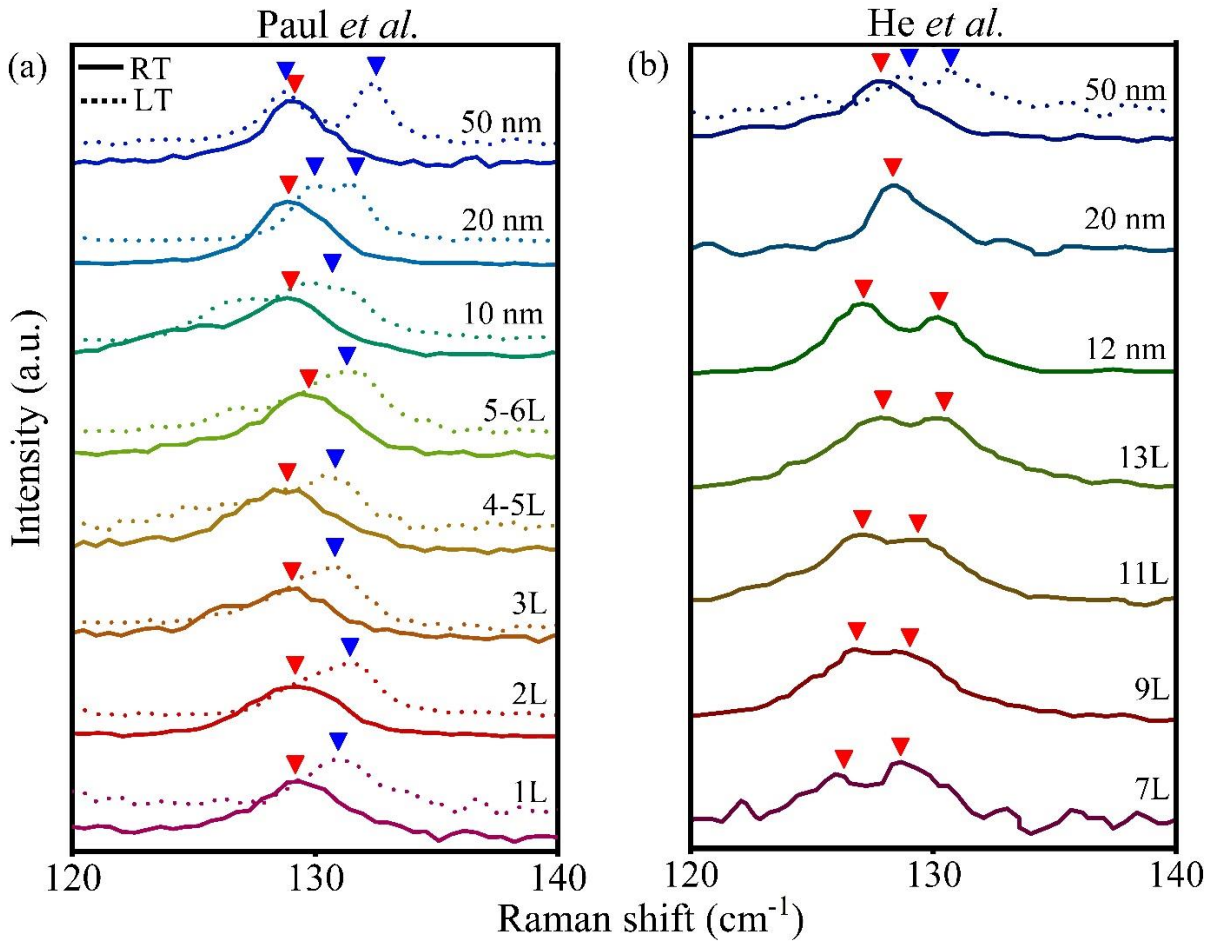

**Figure 8.** Effect of dimensionality on the phase transition. **a)** The comparison of Raman spectra taken at high and low temperatures (represented by solid and dotted spectra) shows a suppression of the phase transition for thin flakes, below the thickness of 20 nm. The thin flakes stabilize in the 1$T'$ phase, as represented by a single Raman active phonon near ~129 cm$^{-1}$. **b)** Similar suppression of phase transition observed with the thin flakes stabilizing in the $T_d$ phase, as represented by the presence of two Raman active modes near ~129 cm$^{-1}$. The Raman active phonons in both **(a)** and **(b)** are marked by the red and blue triangles for the room temperature and low temperature data, respectively. Figure **(a)** is adapted with permission.[32] Copyright 2020, American Physical Soceity. Figure **(b)** is adapted with permission.[33] Copyright 2018, American Physical Soceity.

Paul et al.[32] have studied the thickness evolution of the phase transition through Raman measurements. In this regard, we may recall that the 1$T'$ and the $T_d$ phases are represented by a single Raman active phonon ($P_6$) and two Raman active phonons ($P_{6A}$ and $P_{6B}$), respectively, near the ~129 cm$^{-1}$ region of the Raman spectrum, as discussed previously in section 2. Figure 8a shows the Raman spectra of various flakes of varying thicknesses of MoTe$_2$ obtained at room temperature and at low temperature (80 K). We observe that the flakes show a single Raman active phonon marked with the red triangles (indicating the 1$T'$ phase) at room temperature, for all the investigated flakes, irrespective of their thicknesses. The low temperature Raman spectra for the thick flakes (50 nm and 20 nm) show two Raman active phonons at low temperature (signature of the $T_d$ phase), as expected. But, for thinner flakes, the signature of $T_d$ phase is missing, and all the investigated flakes below 20 nm thickness show signatures of the 1$T'$ phase, even at low temperatures (note the blue triangles to identify the number of Raman active phonons at low temperature for the flakes of various thicknesses). These results suggest that the phase transition is only allowed for thick bulk flakes and is suppressed for thinner flakes of MoTe$_2$. Similarly, He et al.[33] reported the suppression of the phase transition in thin flakes of MoTe$_2$ (Figure 8b). However, though both these reports[32,33] claimed a suppression of the phase transition, the stable phase predicted for the thin flakes was the 1$T'$ phase in Paul et al.'s measurements, while it was $T_d$ phase in He et al.'s work. As a consequence, the Raman spectra for various flakes reported by He et al.[33] show two Raman active phonons (marked with red triangles) up to room temperature. This apparent disparity in the two reports has its roots in the preparation and experimental conditions for the two reports and has been explained by Paul et al.[32]. While He et al. have performed experiments on flakes capped with hBN layer, thereby preventing exposure to atmospheric ambience, the measurements performed by Paul et al. were performed on flakes that were exposed to atmospheric air and moisture that is known to cause hole doping in the system. Paul et al. demonstrated the stabilization of the 1$T'$ phase in thin flakes of MoTe$_2$ that were hole doped and the $T_d$ phase in flakes that were electron doped. This brings us to the discussion of the effect of charge doping on the phase transition.

Before discussing the effect of charge doping, it would be worthwhile to discuss about another important result reported by Paul et al.[32] from their Raman measurements. The temperature-dependent behavior of phonons is predominantly represented by the anharmonic theory which allows the phonons to interact with each other, thereby resulting in temperature-dependent phonon properties. The anharmonic theory predicts a redshift in phonon frequency and an increase in linewidth as a function of temperature[61,62]. This anharmonic behavior explains the temperature behavior of all phonons with only exceptions occurring when the phonons couple with other degrees of freedom like charge or spin. In presence of electron-phonon or spin-phonon couplings, we expect to see deviations from the anharmonic trends[63–65]. The $P_2$ phonon (labelled in Figure 4) shows a complete reversal in characteristics of both the frequency and linewidth as a function temperature (Figure 9a)[32]. As spin-phonon coupling can be rule out in non-magnetic crystals like MoTe$_2$, the anomalous departure in phonon characteristics is attributed to electron-phonon coupling (EPC). Such anomalous temperature-dependent behavior exhibited by the $P_2$ phonon has been previously reported by Zhang et al.[41] and Kuiri et al.[52] which has been attributed to EPC. Further, Zhang et al.[41] reported an asymmetric Fano lineshape (for the phonons $P_7$ ~165 cm$^{-1}$ and $P_{10}$ ~260 cm$^{-1}$), which is again another typical signature of electron-phonon interactions. Paul et al.[32] proposed that the extent of the anomaly (in the temperature-dependent phonon frequency of mode $P_2$) may provide a measure of the strength of the EPC, which in turn is related to the concentration of electrons in the crystal. Figure 9b shows the strength of EPC in MoTe$_2$ flakes of various thicknesses. We observe an increase in EPC with decreasing thickness which is consistent with the observation of electron doping induced by Te deficiencies in thinner flakes of MoTe$_2$[43]. However, in the ultrathin flakes, we observe a sudden decrease in EPC, which finally gets completely suppressed for flakes of monolayer MoTe$_2$. This is because of strong hole doping in the thin flakes due to exposure to atmospheric moisture and air[66,67]. The hole doping induced by the atmospheric moisture and air is a surface phenomenon and is, therefore, effective only for very thin flakes which offers a large surface to volume ratio. Thus, the evolution of EPC with thickness also confirms the fact that the thin flakes of MoTe$_2$ are hole doped. As also discussed previously, the thin layers which are exposed to ambient atmospheric conditions (and are, therefore, strongly hole doped) show the $1T'$ phase as the stable phase. Further, Paul et al.[32] demonstrated that it is possible to stabilize the $T_d$ phase in the thin films of MoTe$_2$ by exposing the flakes to ammonia vapor, which has been reported to induce electron doping (Figure 9c).

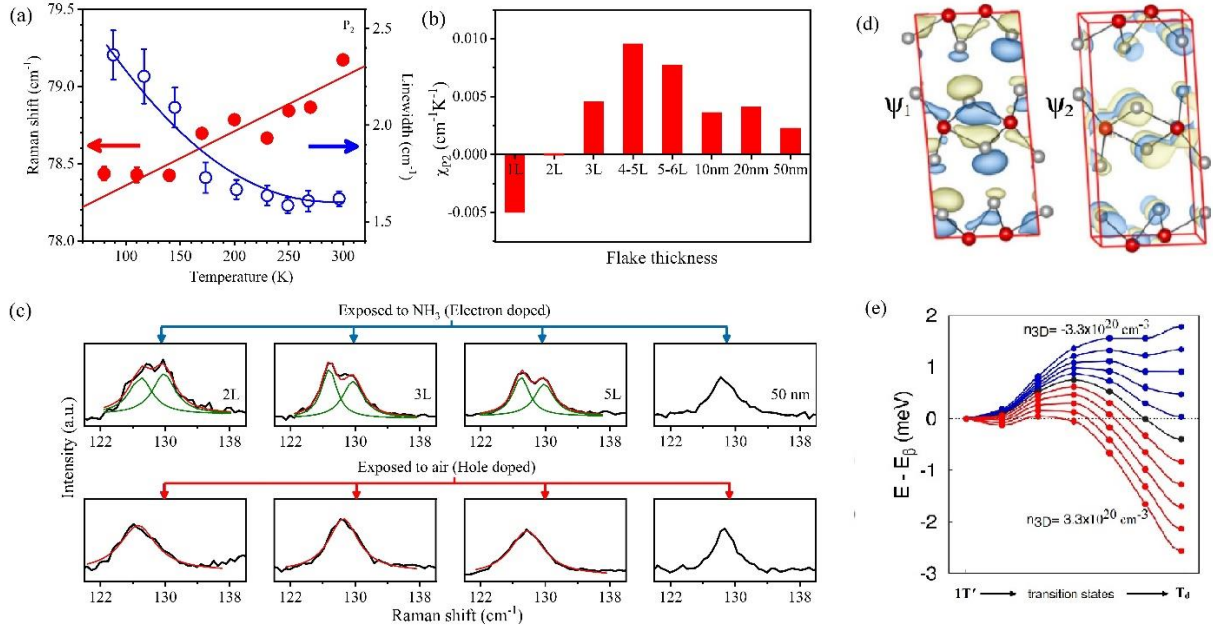

**Figure 9.** Effect of charge doping on the phase transition. **a)** The anomalous behavior of phonon P$_2$ shown in Figure 4a representing electron-phonon coupling (EPC). **b)** The strength of electron-phonon coupling estimated from the extent of anomaly in the P$_2$ phonon behavior, as a function of thickness. The evolution of EPC indicates strong hole doping in thin flakes of MoTe$_2$. **c)** The stabilization of the $T_d$ phase and $1T'$ phase in the electron-doped and hole-doped flakes, respectively. **d)** $\Psi_1$ and $\Psi_2$ wave functions of $1T'$- MoTe$_2$. **e)** Calculated energy profile along the transition path from $1T'$ to $T_d$ phase of MoTe$_2$ as a function of doping ($n_{3D}$). The energy profiles with electron (positive) doping, hole (negative) doping, and neutral case are drawn by red, blue, and black lines, respectively. Figures **(a-c)** are reproduced with permission.[32] Copyright 2020, American Physical Soceity. **(d,e)** are reproduced with permission.[68] Copyright 2017, American Physical Soceity.

Kim et al.[68] came up with a theoretical explanation to the stability of the $1T'$ phase due to hole doping and $T_d$ phase due to electron doping. Through their *ab initio* calculations, they have computed the electronic band structure for both the phases. In the vicinity of the Fermi level, they have spotted the two topmost partially filled valence bands formed by antibonding states between the p and d orbitals of the Te and Mo atoms, respectively. These states are represented by $\Psi_1$ and $\Psi_2$, respectively, and are drawn in Figure 9d to visualize their bonding nature at the $\Gamma$ point. While the $T_d$ phase was found to possess almost a similar electronic band structure to the $1T'$ phase, the antibonding $\Psi_1$ and $\Psi_2$ bands were shifted to lower energies, implying that they were populated more with electrons as the phase transition from the $1T'$ to the $T_d$ phase occurred. This means the $T_d$ phase gets stabilized by addition of more electrons to the antibonding $\Psi_1$ and $\Psi_2$ bands. A similar effect may, therefore, be obtained by doping MoTe$_2$ with electrons. Therefore, they went on to explore the effect of charge doping in the system by studying the energy profile along the phase transition pathway from the $1T'$ to the $T_d$ phase at various concentrations of hole and electron doping (Figure 9e). They observed that while hole doping stabilizes the $1T'$ structure, electron doping stabilized the $T_d$ structure, which was also demonstrated experimentally by Paul et al.[32] for thin flakes of MoTe$_2$, which undergoes suppression of phase transition.

In addition to the dependence of the strength of EPC on charge doping, it has been reported to show sensitivity to sample conditions and disorders that are distinguished by synthesis methods. While MoTe$_2$ crystals prepared by the flux method showed clear signatures of an additional topological transition at ~70 K, the crystals prepared by the chemical vapour transport (CVT) method did not show the additional topological transition. Zhang et al.[41] claimed that this topological transition (~ 70 K) is essentially a transition from topological state characterized by four WPs to a state characterized by two WPs and is driven by strong EPC. The suppression of the topological transition in the samples synthesized by CVT method is attributed to disorders present in the samples that reduce the effect of EPC. On the other hand, Kuiri et al.[52] reported anomalous phonon behaviors below the phase transition temperature ($T_s$) from the $1T'$ to the $T_d$ phase, which they have attributed to strong EPC between the Weyl fermions and the phonons in the WSM $T_d$ phase.

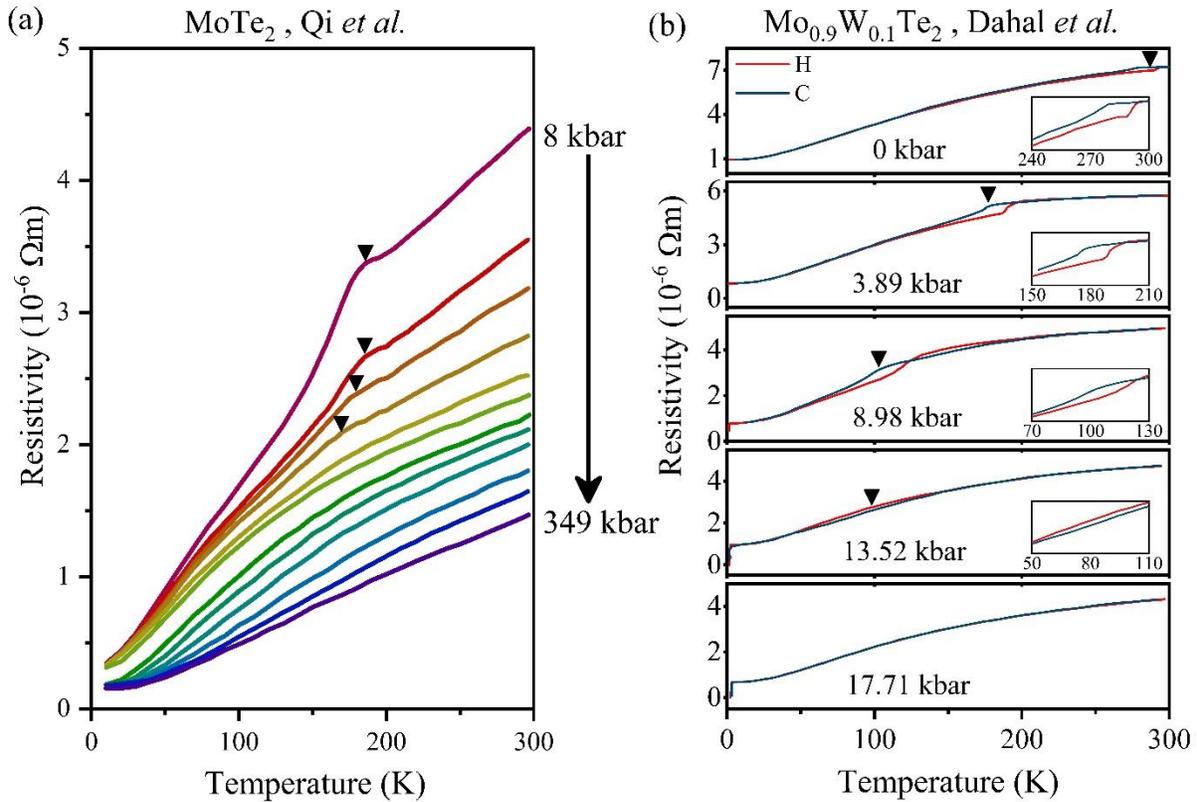

**Figure 10.** Effect of pressure on the phase transition. **a)** Electrical resistivity measurements on MoTe$_2$ as a function of temperature for pressures of 8 kbar to 349 kbar. The anomaly associated with the structural transition is completely suppressed with increasing pressure. **b)** Temperature-dependent resistivity measurements revealing anomalous hysteretic behavior representing the structural phase transition in Mo$_{0.9}$W$_{0.1}$Te$_2$. Insets show zoomed image of the hysteresis loops. Black triangles mark the transition temperature $T_s$ for both **(a)** and **(b)**. Figure **(a)** is adapted from ref. [16] and **(b)** is adapted from ref. [34].

### 5.2. Effect of pressure

The application of external pressure has been reported by Qi et al.[16] to induce an enhancement in the superconducting $T_c$. Similar to the effect of reduced dimensionality, the application of pressure also affects the phase transition from the $1T'$ to the $T_d$ phase. Figure 10a shows the charge transport data by Qi et al.[16] at various pressures, which exhibits the structural transition as an anomaly in the transport data (marked with black triangles) near the structural transition temperature, $T_s$. We observe that the metallic character becomes more pronounced with application of pressure. Further, the extent of the anomaly (which indicates the structural phase transition) and the associated phase transition temperature ($T_s$) reduces with application of pressure. Beyond 40 kbar pressure, the structural phase transition is completely suppressed. This means that the $1T'$ monoclinic phase is stabilized with application of external pressure. This proposition is further confirmed by singlr crystal XRD (SXRD) measurements by Qi et al.[16]. Another recent work by Dahal et al.[34] revealed similar decrease in phase transition temperature and subsequent stabilization of the $1T'$ monoclinic phase by application of pressure on $Mo_{1-x}W_xTe_2$ flakes. Figure 10b shows the data for $Mo_{0.9}W_{0.1}Te_2$ flakes obtained at various pressures. We observe a systematic decrease in the phase transition temperature (indicated by the black triangles pointing the center of the hysteresis loops) with increasing pressure. We also observe a slightly higher $T_s$ (~280 K) for $Mo_{0.9}W_{0.1}Te_2$ at ambient pressure compared to pure $MoTe_2$ crystals, which is an effect of chemical substitution, which we will discuss next.

### 5.3. Effect of chemical substitution

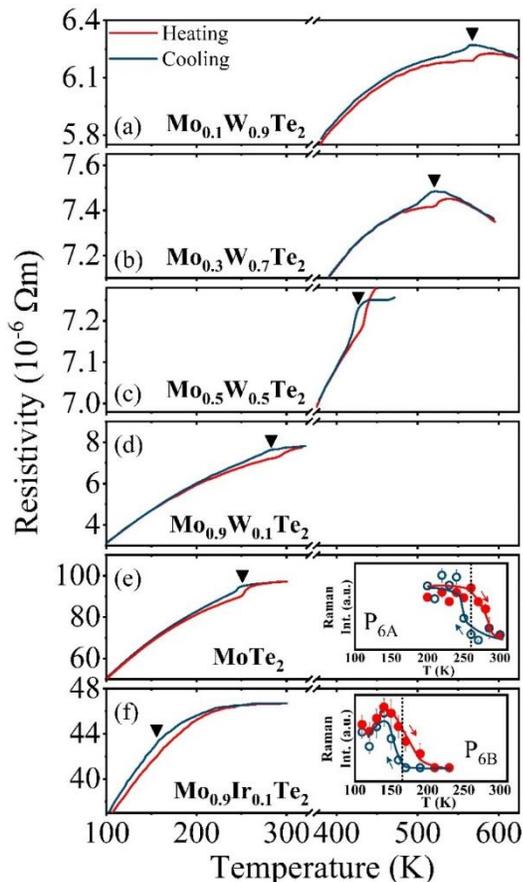

**Figure 11.** Effect of chemical substitution on the phase transition. Temperature-dependent resistivity measurements revealing anomalous hysteretic behavior representing the structural phase transition of $MoTe_2$ is compared for different chemical substitutions in pure $MoTe_2$: **a-d)** W-doping by 90, 70, 50, and 10%, respectively, **e)** pure $MoTe_2$, and **f)** 10% Ir-doping. Insets of **(e)** and **(f)** show the hysteresis loops obtained from Raman measurements, showing same $T_s$ as obtained from transport data. Black triangles mark the transition temperature $T_s$. Figures **(a-d)** are adapted from ref. [34], **(e-f)** are adapted from ref. [36].

Chemical substitution is again another effective means of enhancing the superconducting phase transition temperature[26,32,34–36,49,69] and like the other cases discussed above, also affects the phase transition properties. There are various reports of chemical substitution in $MoTe_2$, some of which have been combined in Figure 11. The figure shows the phase transition of $MoTe_2$ crystals with various chemical substitutions through their anomalous hysteretic charge transport behavior in the vicinity of $T_s$. We observe that 10% Ir doping at the Mo site results in a reduction of the phase transition temperature $T_s$ from ~260 K (Figure 11e) to ~155 K (Figure 11f)[36]. The corresponding Raman signatures of the hysteretic change in phonon properties for pure and 10% Ir doped $MoTe_2$ is shown in the inset of Figure 11e and Figure 11f. On the other hand, doping the Mo site with W by 10%, 50%, 70%, and 90% result in enhancements of the $T_s$ to 277, 423, 505, and 565 K, respectively[34]. The effect of chemical substitution may include a combination of various effects like internal pressure in the lattice, charge doping, etc., which in turn shows variations in the phase transition properties of the material, as discussed above. In order to explain the trend observed in the phase transition temperature, we may invoke the internal pressure generated in the crystal due to variation in ionic radii of the doped atom. The ionic radii of the $Mo^{4+}$, $Ir^{4+}$, and $W^{4+}$ ions are 0.65, 0.625, and 0.66 Å, respectively. It is easy to understand that replacing an ion in a crystal by a smaller ion may result in a compressive pressure on the crystal, which is similar to an externally applied pressure. Therefore, replacing a Mo atom with an Ir atom in the $MoTe_2$ crystal would mimic the effect of an externally applied pressure, while replacement of a Mo atom with W atom would result in an opposite behavior. This is consistent with the fact that Ir doping brings the $T_s$ down to lower temperature (similar to the effect of externally applied pressure discussed previously[16]), and W doping enhances the $T_s$.

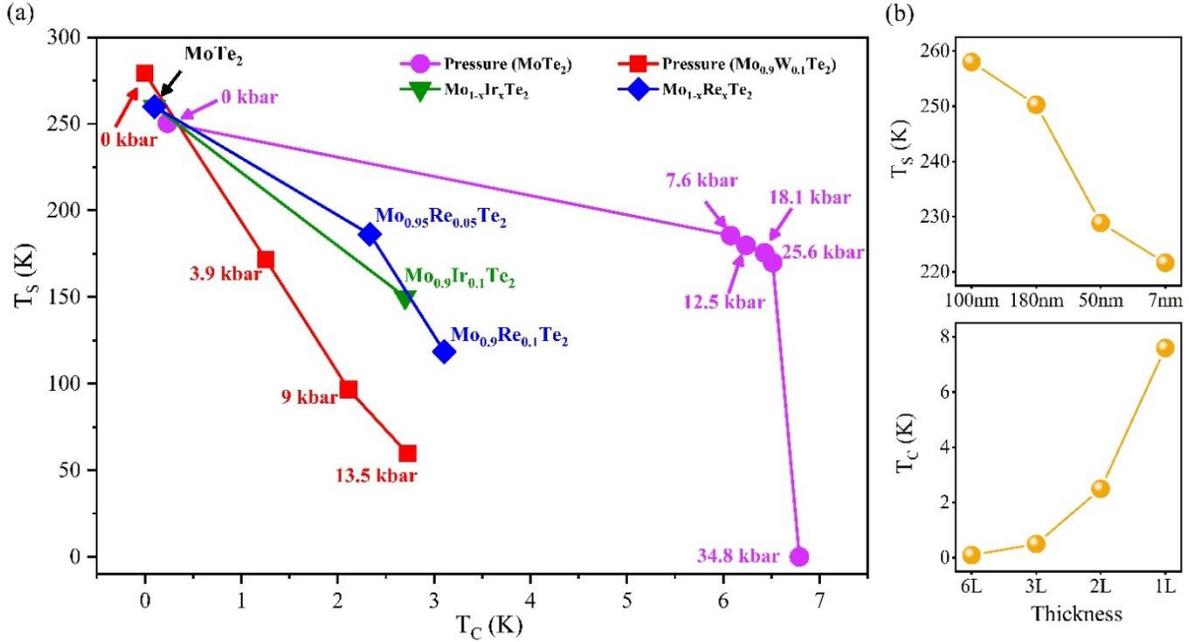

**Figure 12. a)** A correlation between the topological phase transition temperature ($T_s$) and the superconducting critical temperature ($T_c$). The red squares and purple circles represent data points for increasing applied pressure on flakes of pure and 10% W doped MoTe2, the olive inverted triangle and the blue rhombuses represent data points corresponding to Ir and Re doping, respectively. In cases where the phase transition is completely suppressed, we have labeled the $T_s$ as 0 K. **b)** The effect of varying thickness on the $T_s$ and $T_c$ of MoTe2. It may be noted that the $T_s$ decreases as a function of decreasing thickness for flakes of MoTe2 down to a thickness of 7 nm. For ultrathin flakes with thicknesses as low as 6L and below, the $T_c$ shows an enhancement with decreasing thickness. However, these thin flakes are supposedly in the $T_d$ phase (high temperature phase which breaks inversion symmetry). The structural transition temperature in such thin flakes get suppressed, but these flakes may stabilize in either of the two structures ($1T'$ or $T_d$) depending on whether the flakes are doped with holes or electrons.

## 6. Possible correlation between the superconducting and topological phase transitions

As already discussed, semimetallic MoTe2 undergoes a superconducting phase transition at very low temperature. Recalling the phase transition temperature ($T_s$) from the $1T'$ phase to the $T_d$ phase for bulk MoTe2 at $T_s = 250$ K, we may conclude that the $T_d$ phase (which is also a Type-II WSM) hosts a superconducting ground state with a $T_c$ of ~0.10 K[16,26–31]. However, the ultralow $T_c$ for the superconducting transition renders its usage and realization extremely unlikely. Therefore, numerous attempts to induce the superconducting transition at higher temperatures have been proposed and demonstrated through the application of stimuli like reduced dimensionality, pressure, charge doping, chemical substitution, etc. In the previous section, we have elaborately discussed how these stimuli also influence the topological phase transition. Based on various reports, a possible correlation between the topological and superconducting transitions may be noted. Figure 12a shows a correlation map which plots the $T_s$ against $T_c$ of bulk samples of MoTe2 studied as a function of various parameters. For example, the effect of i) applied pressure on flakes of pure[16] as well as 10% W doped MoTe2[34] and ii) chemical substitution (Ir[36] and Re[35] doping) on the $T_s$ and the $T_c$ values have been plotted. It may be noted that with increasing pressure and increasing amount of chemical substitution (with ions of smaller ionic radii compared to $Mo^{4+}$), the $T_c$ shows an increment with a simultaneous decrement in the topological phase transition temperature or its complete suppression. In cases where the phase transition is completely suppressed, we have labeled the $T_s$ as 0 K, in Figure 12a. This means, while we expect to observe the superconducting ground state in the $T_d$ phase (which is topological in nature), an enhancement of the $T_c$ might result in the superconducting ground state to be present in the $1T'$ phase, which has been reported to be topologically trivial[16]. Though there is no available physical explanation for the appearance of a correlation (between the topological phase transition and the superconducting ground state) or its existence, we believe the topic may be extremely important to establish the possible link between the two apparently unrelated phenomena in MoTe$_2$. Figure 12b shows the effect of reduced dimensionality on the $T_s$ and $T_c$ of MoTe$_2$. Following a trend very similar to the other perturbative effects like pressure and chemical substitution (shown in Figure 12a), thinning down the flake thickness also shows a decrease in $T_s$[33]. However, this trend is observed for thicknesses down to 7 nm. For thinner flakes, as discussed previously, the structural transition is absent. But such thin flakes might exhibit either of the two structures, depending on the type of inherent charge doping in the sample. For the hole doped samples, the $1T'$ phase gets stabilized, while the $T_d$ phase stabilizes for the electron doped samples. Consequently, the structural transition temperature, $T_s$ is an ill-defined quantity for such thin flakes of MoTe$_2$. Nevertheless, it can be observed that the superconducting $T_c$ shows an enhancement with decreasing flake thickness[29] in the ultrathin flakes of MoTe$_2$ (Figure 12b, lower panel). Therefore, unlike bulk crystals, where a correlation between the $T_s$ and $T_c$ of MoTe$_2$ apparently exists, it may be possible to obtain superconducting ground states in the thin flakes at elevated temperatures in both the centrosymmetric $1T'$ phase or the topological $T_d$ phase, with suitable charge doping. However, further studies (both theoretical and experimental) are required to better understand such possibilities.

## 7. Outlook

Despite the rapidly flourishing research on the topological phase transition in MoTe$_2$ in recent years, we have noted an obvious amount of ambiguity and lack of understanding in certain areas. In this section, we will pose some open questions that demand the attention of the theoretical and experimental researchers alike and may serve as interesting research areas in the immediate future. Firstly, though the association of the topological phase transition with a structural change is well established, the phase transition pathways are not well understood. The recent reports on the appearance of an intermediate $T_d^*$ phase (while undergoing transition from the $T_d$ to the $1T'$ phase) through neutron scattering experiments needs further confirmations via other experimental means. Besides, the reports have shown the appearance of the intermediate phase ($T_d^*$) in the heating cycle[38,40], but it could not be observed in the cooling cycle. This further renders the present understanding of the phase transition pathways as inadequate. Secondly, the observation of a simultaneous decrease in the phase transition temperature ($T_s$) and an increment in the superconducting transition temperature ($T_c$) in bulk crystals of MoTe$_2$, when perturbed by various stimuli like pressure and chemical substitution, invokes the possibility of a correlation between these two apparently unrelated phenomena. Finally, the effect of dimensionality brings out even more complications as we observe that for flakes of thicknesses below a critical value, the structural transition is suppressed with the possibility to stabilize either the $1T'$ or the $T_d$ structure at all temperatures by suitable charge doping. Such thin flakes have also shown an increment in the superconducting $T_c$, which imply that thin flakes may likely show the superconducting transition in the desired phase, with suitable charge doping. These propositions currently have no theoretical basis and the observations may be merely coincidental. Therefore, an extensive amount of research is still required to better understand the phase transition and possibly answer some of the open questions mentioned above. This may also lead to exciting new applications associated with topological transitions, superconductivity, and a possible route to control these phenomena through correlations between these properties.

## 8. Conclusions

We have reviewed the topological phase transition between the $1T'$ and $T_d$ phases of MoTe$_2$. We have discussed in detail about the various changes in the structural and electronic properties of the MoTe$_2$ crystal that occur around the phase transition temperature, and the possible detection techniques of the transition through the study of associated phenomena. We have then, went on to study the effect of various external stimuli like dimensionality, charge doping, pressure, chemical substitution etc. on the phase transition. As these stimuli have also been reported to show enhancements in the superconducting T$_c$ of MoTe$_2$, we have tried to establish the presence of a correlation between the topological and superconducting transitions in MoTe$_2$, by comparing data from various reports.

**Acknowledgements**

The authors acknowledge funding from the Department of Science and Technology, Science and Engineering Research Board (Grants No. ECR/2016/001376 and No. CRG/2019/002668) Ministry of Education (Grant No. STARS/APR2019/PS/662/FS) and Nanomission [Grant No. SR/NM/NS-84/2016(C)]. The authors also acknowledge Mr. Arpan Ghosh (BSMS student, IISER Bhopal) for his contributions in the preparation of Figure 1.

**Keywords:** Topological phase transition, Symmetry breaking, Centrosymmetric crystal, Quantum spin Hall effect, Weyl semimetal, Fermi arc, Perturbations.

**Authors:**

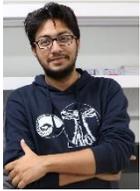
**Suvodeep Paul** received his M.Sc. from Jadavpur University, Kolkata. He is currently a Senior Research Fellow at the Department of Physics, Indian Institute of Science Education and Research Bhopal. His research interests are novel two-dimensional materials and understanding their electronic and phonon properties.

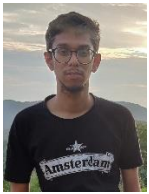
**Saswata Talukdar** received his M.Sc. from the University of Calcutta, Kolkata. He is currently a Junior Research Fellow at the Department of Physics, Indian Institute of Science Education and Research Bhopal. His research interests are novel two-dimensional materials and understanding their electronic and phonon properties.

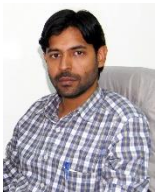
**Ravi Shankar Singh** received his Ph.D. from Tata Institute of Fundamental Research, Mumbai. He is currently an Associate Professor at the Department of Physics, Indian Institute of Science Education and Research Bhopal. His research interests are correlated condensed matter systems, photoemission spectroscopy, and Density Functional Theoretical calculations.

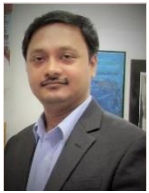
**Surajit Saha** received his Ph.D. from Indian Institute of Science, Bangalore. He is currently an Associate Professor at the Department of Physics, Indian Institute of Science Education and Research Bhopal. His research interests are correlated condensed matter systems and Raman scattering at extreme conditions.